\documentclass[aps,prl,reprint]{revtex4-1}
\usepackage{blindtext}
\usepackage{graphicx}
\usepackage{graphics}
\usepackage{bm}        
\usepackage{verbatim}   
\usepackage{amsfonts}
\usepackage{amsmath}
\usepackage{bm}
\usepackage{amssymb}
\usepackage{subfigure}
\usepackage{adjustbox}
\usepackage{lipsum}

\usepackage{adjustbox}

\usepackage{stfloats}

\usepackage{amsmath,amscd}
\usepackage {extarrows}
\usepackage{algorithm,algpseudocode}
	
\makeindex

\begin{document}

\title {Non Gaussian information of heterogeneity in Soft Matter}

\author{Rahul Dandekar $^a$, Soumyakanti Bose $^b$ and Suman Dutta $^{\ast,a}$ }

\affiliation{$^{a}$The Institute of Mathematical Sciences, \\  IV Cross Street, CIT Campus, Taramani, Chennai,\\ Tamilnadu 600113.\\
$^{b}$ Indian Institute of Science Education and Research, Knowledge city, Sector 81, Manauli PO, Sahibzada Ajit Singh Nagar, Punjab 140306}

\email{RD and SB contributed equally. Corresponding author: sumand@imsc.res.in}

\date{~\today}

\begin{abstract}
Heterogeneity in dynamics in the form of non-Gaussian molecular displacement distributions appears ubiquitously in soft matter. We address the quantification of such heterogeneity using an information-theoretic measure of the distance between the actual displacement distribution and its nearest Gaussian estimation. We explore the usefulness of this measure in two generic scenarios of random walkers in heterogeneous media. We show that our proposed measure leads to a better quantification of non-Gaussianity than the conventional ones based on moment ratios.\\
\end{abstract}


\maketitle

The usual laws of Fickian diffusion have dominated the world of molecular transport for more than a century \cite{dh1}. Fickian diffusion models the evolving probability distribution of the molecular displacement in a medium as a Gaussian when there is an extreme separation of time scales between the molecular solute and the atomistic solvent particles \cite{dh1,sm1}. Since the relaxation of atomistic solvent particles is many magnitudes faster than the larger solute particles, the solute dynamics is treated as a molecular dynamics with effective fluctuations due to constant thermal kicks by the bath particles. However, this is not \emph{a priori} valid locally in systems with rough energy landscapes and multiple time-scales of relaxation \cite{dh1,ng-bouchad1}.

In the presence of multiple relaxation time-scales, the molecular displacement distributions do deviate from Gaussianity\cite{ng-bouchad1,ng-bouchad2,dh2, ng-pc,ng-granick1,ng-granick2,ng-chubensky}. This is observed ubiquitously in soft matter\cite{ng-bouchad1, ng-pc}, from soft glasses to bio-molecules \cite{ng-bouchad2, ng-granick1}, even if the dynamics is essentially bounded by the central limit theorem and thus the distribution reverts to a Gaussian form at timescales much larger than those of molecular relaxation\cite{ng-granick2,dutta1,jain-seb, ng-metzler1,ng-metzler2}. Also, in presence of domains, effective polydispersity can give rise to heterogeneity  in dynamics where each of the molecular hops from one domain to another corresponds to unique time scales while the intra-domain movements mostly mimic the dynamics in bulk\cite{dutta3}. 
Similar heterogeneity is seen in glassy liquids due to intermittency caused by local metastability, where motion in the vicinity of a local cage becomes arrested, and becomes diffusive again when it gets out of the cage overcoming the free energy barrier\cite{cage-sk,cage-pc}.
Thus the coexistence of competing relaxation processes leads to dynamic heterogeneity in soft matter\cite{cage-sk,cage-pc}. This necessarily asks for validation and generalisation of Einstein-Stokes in many physical systems where relaxation processes undergo heterogeneity.

One phenomenological approach to understand non-Gaussian diffusion is to consider that the effective single particle dynamics is characterized by a diffusion spectrum, $P(D)$, instead of a unique diffusion coefficient $D$\cite{ng-granick2}. Non-Gaussian probability distributions of particle displacements are thus modeled as an ensemble of diffusive processes, $G(x,t;D)$, given by $P_{\rm{ng}}(x,t)\sim \int dD~ P(D)~ G(x,t;D)$ \cite{ng-granick2,ng-chubensky}. Here the weighted distribution, $P(D)$ captures the dynamical fluctuations in molecular diffusion. In fact, this picture of {\em Fickian yet Non-Gaussian} diffusion has been quite successful in explaining some natural processes which  earlier seemed anomalous\cite{ng-granick2,ng-chubensky,dutta1,jain-seb}. From here, one ends with a surprising conclusion is that the linearity of the mean squared displacements no longer ensures the Gausianity of the underlying probability distribution function\cite{ng-granick2}. Thus, a system with a diffusivity spectrum, $P(D)$ can yield $P_{\rm{ng}}(x,t)\sim \exp(-\frac{x}{\sqrt{t}})$, leading to a linear MSD: $\langle x^{2} \rangle=D_{\rm{eff}}t = \int dD P(D) D$, the effective distribution of particle displacements $P_{\rm{ng}}(x,t)$  remains non-Gaussian. Thus the actual distribution of particle displacements contains information about the heterogeneity which is masked if one approximates the diffusion by the Gaussian approximation, $P_{\rm{eff}}(x,t) \approx P^{g}_{ng}(x,t) \sim \frac{1}{\sqrt{tD_{\rm{eff}}}} e^{-x^2/4 D_{\rm{eff}} t}$.

In this letter, we introduce a new information-theoretic quantification to tackle the immense challenge of extracting the heterogeneity in molecular diffusion. We discuss an approach to detect such heterogeneity in the form of non-Gaussian information more efficiently than conventional approaches. This is difficult because the information needs to be separated from the expected increase in information due to diffusion. This expected increase is that which happens even for simple random walks, as the intrinsic Shannon information even in a spreading Gaussian increase logarithmically with elapsed time. We compare two quantities in extracting this non-trivial information. We explicitly show that the new quantification we propose leads to more efficient detection of the true non-Gaussian information than what the standard non-Gaussian parameter captures. We exemplify this in two generalised cases where the existence of dynamic heterogeneity is well known and is known to appear due to the intrinsic structural heterogeneity of the medium.   

The conventional approach of quantifying non-Gaussianity in a distribution considers deviation from moment relationships which hold for a Gaussian distribution. The standard {\it Non Gaussian Parameter} is defined as

\begin{equation}
\alpha_2 = \frac{\langle x^4 \rangle}{3 \langle x^2 \rangle^2} - 1 \mbox{ (in 1D) } \frac{3 \langle r^4 \rangle}{5 \langle r^2 \rangle^2} -1 \mbox{ (in 3D)}. \label{ngdef}
\end{equation}

This relies on the fact that for a 1D Gaussian, $\langle x^4 \rangle = 3 \langle x^2 \rangle^2$, and for a 3D Gaussian, $\langle r^4 \rangle = \frac{5}{3} \langle r^2 \rangle^2$, where $r$ is the radial coordinate and $<x^{n}>=\int dx x^{n} P(x)$. 

However, the reliance of this quantity on the fourth moment necessarily limits the amount of information it captures. Quantifications involving higher moments can also be defined, but there is no sensible way to integrate the information in these separate quantities to form a unified picture of non-Gaussianity.

\emph{Non-gaussianity from relative entropy:} We now define our new information theoretic quantification to capture the similarity between the given function and it's closest Gaussian. The definition is based on Kullback-Leibler (KL) divergence \cite{kullback} between the given non-Gaussian probability distributions, $P_{ng}$ and it's Gaussian counterpart, $P^{g}_{ng}$ having the same first two moments same as $P_{ng}$, 

\begin{equation}
D_{\rm{KL}}(P_{ng}||P^{g}_{ng})=-\int dx P_{ng}(x) \ln \left[ \frac{P^{g}_{ng}(x)}{P_{ng}(x)} \right]\label{def-dkl}
\end{equation}

When quantifying non-Gaussianity, the information gain over the best Gaussian estimate, a simplification occurs,

\begin{equation}
   \Delta S_{\rm{gain}}=D_{KL}(P_{\rm{ng}}||P_{\rm{ng}}^{\rm{g}}) =S_{ng}^{g} - S_{ng} \label{def-re}
\end{equation} 
	
where $S_{ng}=S[P_{ng}]$, $S^{g}_{ng}=S[P^{g}_{ng}]$ and $S(P(x))=\int dx \frac{1}{\sqrt{2\pi}}P(x)\ln{P(x)}$, the Shannon entropies \cite{shannon} of the distributions. Thus, for a Gaussian distribution,$P_{G}$, $ \Delta S_{\rm{gain}}=D_{KL}(P_{G}||P^{g}_{G})=0$. This quantification has earlier been proposed in analysing non-Gaussianity in quantum optical states \cite{iks}. This is also sometimes referred as {\it Negentropy}.

{\it Fickian yet Non-Gaussian Random walk:} We first consider a typical case of molecular heterogeneity adopting the phenomenological picture discussed in Ref.\cite{ng-granick2} when the single particle dynamics is estimated as a convolution of diffusive processes with the distribution of diffusivities, $\tilde{P}(D)$ instead of a single diffusion coefficient:

\begin{equation}
P_{\rm{ng}}(x,t) = \int dD G(x,t;D) \tilde{P}(D)    \label{ng-def}
\end{equation}

where the normal diffusive process be given by $G(x,t;D) (\sim \frac{e^{-\frac{x^{2}}{Dt}}}{\sqrt{2\pi Dt}})$. The peculiarity of this kind of diffusion is that, it leads to the linear mean-squared displacements as in normal diffusion. But in contrast to normal diffusion, it has a diffusion spectrum to represent it's single particle dynamics. Since, diffusion is generally found finite and range bounded ($D\in [D_{0},D_{l}]$) in physical systems, we consider, $P_{\rm{ng}}(x,t)=\int_{D_{0}}^{D_{l}} dD G(x,t;D) \tilde{P}(D)$.

 For such a form of $P_{ng}(x,t)$, using expression of Eq. \eqref{ng-def} in Eq. \eqref{ngdef}, we obtain, 

\begin{equation}
    \alpha_{2}=\frac{\langle D^{2} \rangle - \langle D \rangle^{2}}{\langle D \rangle^{2}}=\frac{\Delta D^{2}}{\langle D \rangle^{2}} \label{alpha_sec1}
\end{equation}

where $\Delta D^{2}=\langle D^{2} \rangle- \langle D \rangle^{2}$. One can check that for $P(D)\sim \delta (D-D_{eff})$, we get $\alpha_{2}=0$. 

 The relative entropy based quantification in Eq. \eqref{def-dkl} and \eqref{def-re} uses {\em statistical distance} of $P_{\rm{ng}}(x,t)$ from its best Gaussian approximation $P_{\rm{ng}}^{\rm{g}}(x,t)$, which is given by
\begin{equation}
P_{\rm{ng}}^{\rm{g}} = \frac{1}{2\pi \langle D \rangle t} e^{-\frac{x^2}{4 \langle D \rangle t}}. \label{bestgauss}
\end{equation}
 
This quantifies the non-Gaussianity of $P_{\rm{ng}}(x,t)$ also, in turn, provides the {\em gain in information} due to heterogeneity, $\Delta S_{gain}$. Using the concavity of the logarithm, i.e., $\ln \left[ \int_{D_{0}}^{D_{l}} dD G(x,t;D) \tilde{P}(D) \right] \geq \int_{D_{0}}^{D_{l}} dD  \tilde{P}(D) \ln [G(x,t;D)]$\cite{cover}, we obtain 

\begin{equation}
     S_{\rm{ng}}(t) = \left\langle D \right\rangle \left\langle \frac{1}{D} \right\rangle + \frac{1}{2} \left( -1 + \ln [2\pi] + \left\langle \ln [Dt] \right\rangle \right) - F_{1}(D),
\label{eq_exp_sng}
\end{equation}
where $F_1(D)$ is a strictly positive quantity and only a function of $P(D)$. We have also used the fact that
\begin{equation}
- \int_{C} dx G(x,t;D) \ln [G(x,t;D)] = \frac{1}{2} (1 + \ln [2\pi Dt]).
\end{equation}
The origin of $F_{1}(D)$ lies in the convex combination of different local diffusive processes.  As the heterogeneity increases, $F(D)$ becomes larger indicating higher departure from the unique diffusivity.

On the other hand the entropy of the best Gaussian approximation, eqn. (\ref{bestgauss}), is given by
\begin{align}
S_{\rm{ng}}^{\rm{g}}(t)&=-\int_{C} dx P_{\rm{ng}}^{\rm{g}}(x,t) \ln \left[ P_{\rm{ng}}^{\rm{g}}(x,t) \right] = \frac{1}{2} ( 1 + \ln [2\pi \langle Dt \rangle ])
\nonumber
\\
& \geq   \frac{1}{2} ( 1 + \ln [2\pi] + \langle \ln [Dt] \rangle 
\nonumber
\\
&= \frac{1}{2} ( 1 + \ln [2\pi] + \langle \ln [Dt] \rangle  + F_{2}(D).
\label{eq_exp_sngg}
\end{align}

Similar to $F_{1}(D)$, $F_{2}(D)$ in Eq. (\ref{eq_exp_sngg}) also arises due to the convolution over all the local Gaussian distributions. 
However, unlike the previous case, it only yields the excess entropic contribution due to the Gaussian approximation.

Putting the results from Eq. (\ref{eq_exp_sng}) and Eq. (\ref{eq_exp_sngg}) in Eq. (\ref{def-re}) we get 
\begin{align}
\Delta S_{\rm{gain}} &= F_{2}(D) + F_{1}(D) + 1 - \left\langle D \right\rangle \left\langle \frac{1}{D} \right\rangle
\nonumber
\\
&= F_{1}(D) + F_{2}(D) - \left( \left\langle D \right\rangle \left\langle \frac{1}{D} \right\rangle -1 \right),
\label{eq_exp_infgain}
\end{align}
as by Jensen's inequality \cite{cover} $\left\langle \frac{1}{D} \right\rangle \geq \frac{1}{\left\langle D \right\rangle}$ for any bonafide distribution $P(D)$.
The positivity of $\Delta S_{\rm{gain}}$, ensured by the positivity of KL divergence, sheds some light on the functional dependence between $F_{1}(D)$ and $F_{2}(D)$. However, the exact form of these quantities depends on the details of the system under consideration.

As an example, we now consider a typical class of non-Gaussian yet Fickian Random walk which is represented by a diffusion spectrum $~P(D;\lambda)$ with a parameter $\lambda$ that controls the heterogeneity in the system. We take a form of $P(D;\lambda)$ such that $\frac{\partial \ln P(D)}{\partial D}$ remains a constant\cite{ng-granick2}: 

\begin{equation}
P(D;\lambda) = \lambda e^{-\lambda (D-1)} \mbox{ for } D>1 \label{eq:expPD}
\end{equation}

The displacement distribution for such a random walker after a time $t$ is given by$P_{ng}(x,t;\lambda) = \int_{1}^{\infty} \frac{P(D;\lambda)}{\sqrt{4\pi D t}} e^{-\frac{x^2}{4 D t}}$. The best Gaussian approximation to this distribution at time $t$ is the one with the same variance as $P_{ng}(x,t;\lambda)$, viz. $2 \langle D \rangle t =2 (1+\lambda^{-1})$. We denote this distribution by $P_{ng}^g(x,t;\lambda)$. For the $P(D)$ given in eqn. \eqref{eq:expPD}, we can calculate a closed form for $P_{ng}(x,t;\lambda)$:

\begin{align}
P_{ng}(x,t;\lambda) &= \frac{\sqrt{\lambda} e^{\lambda}}{4\sqrt{t}} \Big( e^{-|x|\sqrt{\frac{\lambda}{t}}} \rm{Erfc} \big( \sqrt{\lambda} - \frac{|x|}{\sqrt{4t}} \big) + \nonumber \\
&~~~~~~~~~~~~~~~~~e^{|x|\sqrt{\frac{\lambda}{t}}} \rm{Erfc} \big( \sqrt{\lambda} + \frac{|x|}{\sqrt{4t}} \big) \Big) \label{png1}
\end{align} 

Where $\rm{Erfc}(x)$ is the complementary Error function. We use $P_{ng}(x,t;\lambda)$ as in Eq.\eqref{png1} to obtain the non Gaussian parameter:

\begin{equation}
\alpha_2 = \frac{\lambda^{-2}}{\lambda^{-2} + 2\lambda^{-1} + 1}
\end{equation}

Using the form of $P(x,t;\lambda)$ derived above in Eq. 11, we compute $\Delta S_{\rm{gain}}$ by numerical integration with the suitable construction of instantaneous $P^{g}_{ng}(x,t,\lambda)$ that has the same first two moments equal to the one of $P_{ng}(x,t;\lambda)$, following Eq. \eqref{def-re}. 

We now compare the conventional measure of non-Gaussianity with our measure based on the KL divergence, see Fig. 1(a). 
The two measures, $\alpha_{2}$ and $\Delta S$, show a non linear correlation. We see that although $\alpha_2$ saturates for high values of $\lambda^{-1} = \langle D \rangle -1$, $\Delta S_{\rm{gain}}$ does not. We thus see that $\Delta S$ represents more information about the non-Gaussianity than $\alpha_2$, and can distinguish between cases where $\alpha_2$ might not differ appreciably.


\begin{figure}[h]
	\centering
	\subfigure[]{
	\includegraphics[width=0.8\linewidth]{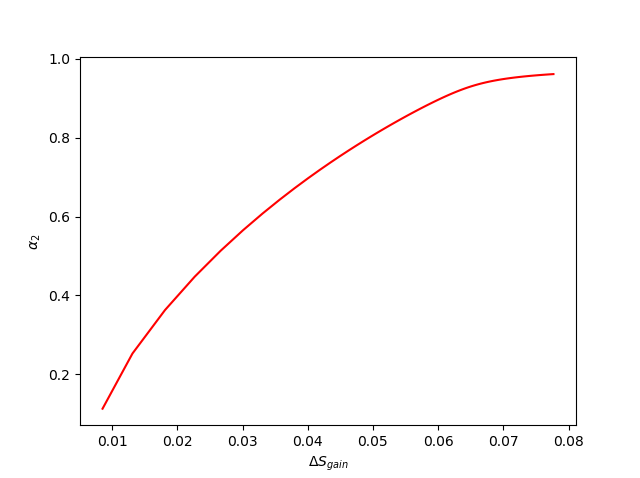}
	}
	
	\subfigure[]{
	\includegraphics[width=0.8\linewidth]{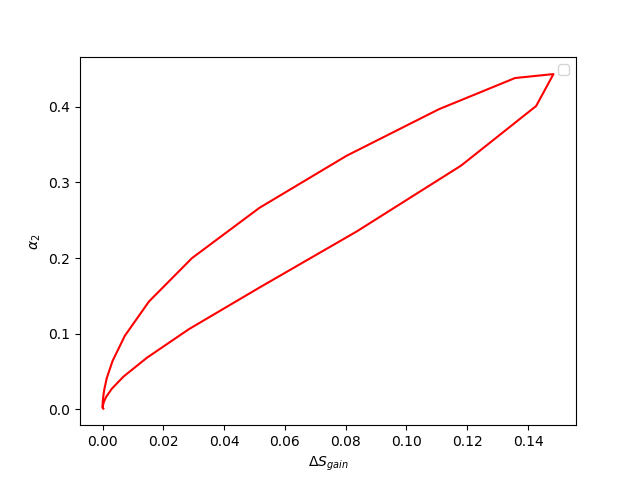}
	}
	\caption{(a) Comparing the information theoretic quantification ($\Delta S$) with the conventional measure ($\alpha_{2}$) for the case of (a)  {\it Fickian yet Non Gaussian Random walk} whose dynamics is governed by Eqn. (10) (b) a walker in a super-cooled liquid following Ref.\cite{langer}.}
	\label{glassywalk}
\end{figure}

The diffusions considered so far can all be expressed as superpositions of Gaussian diffusions with a distribution of diffusion constants, $P(D)$. For this case, it is easy to see that the measure of non-Gaussianity $\Delta S_{\rm{gain}}$ does not change with time. This is because the time-dependent parts of both $S_{ng}$ and $S_{ng}^g$ are equal and equal to $\frac{d}{2}\log{(t)}$, where $d$ is the number of dimensions.

{\it Difffusion in a super-cooled Liquid:} In order to validate this non linear dependence of the two quantifications of non-Gaussianity, we now switch to a more realistic model of heterogeneous system where we look at the transient development of heterogeneity and how its disappearance. The model was proposed by Langer and Mukhopadhyay \cite{langer} in the context of the diffusion of tracer particles in a supercooled liquid, with distinct domains which can be glassy or non-glassy. When in a glassy domain, the particle does not move at all, and only moves once the walls of the domain, which diffuse at a rate $1$, cross the position of the walker, thus leaving it in an non-glassy (or \textit{mobile}) domain. The waiting time of the walker, assuming a distribution of glassy domain sizes $W(\rho) \sim \rho^2 \exp{(-\rho^2)}$, comes out to be $\psi_G(t) = 2 e^{-2 \sqrt{t}}$. In a mobile region, the particle is free to diffuse, and it diffuses with a diffusion constant $\Delta$, which is assumed to be larger than $1$. It now diffuses until it crosses a domain wall into a glassy domain. The distribution of waiting times in a mobile domain is $\psi_M(t) = \Delta e^{-\Delta t}$.

If the walker begins in an unfrozen domain, and assuming spherical symmetry (thus integrating over the angles), one can use the theory of continuous-time random walks to find that \cite{langer}

\begin{equation}
\tilde{P_{ng}}(k,u) = \frac{1}{u}\frac{1- \tilde{\psi_G}(u) + u/\Delta}{1- \tilde{\psi_G}(u) + k^2/2 + u/\Delta}
\end{equation}

which is the Fourier- and Laplace-transformed form of the three-dimensional spherically symmetric probability distribution $P_{ng}(r,t)$ of the position of the particle at time $t$, and $\tilde{\psi_G}(u)$ is the Laplace transform of the waiting time distribution in glassy domains. The above equation can be numerically inverted in Fourier and Laplace space to find $P_{ng}(r,t)$.

\begin{figure}[h]
	\centering
	\includegraphics[width=1.0\linewidth]{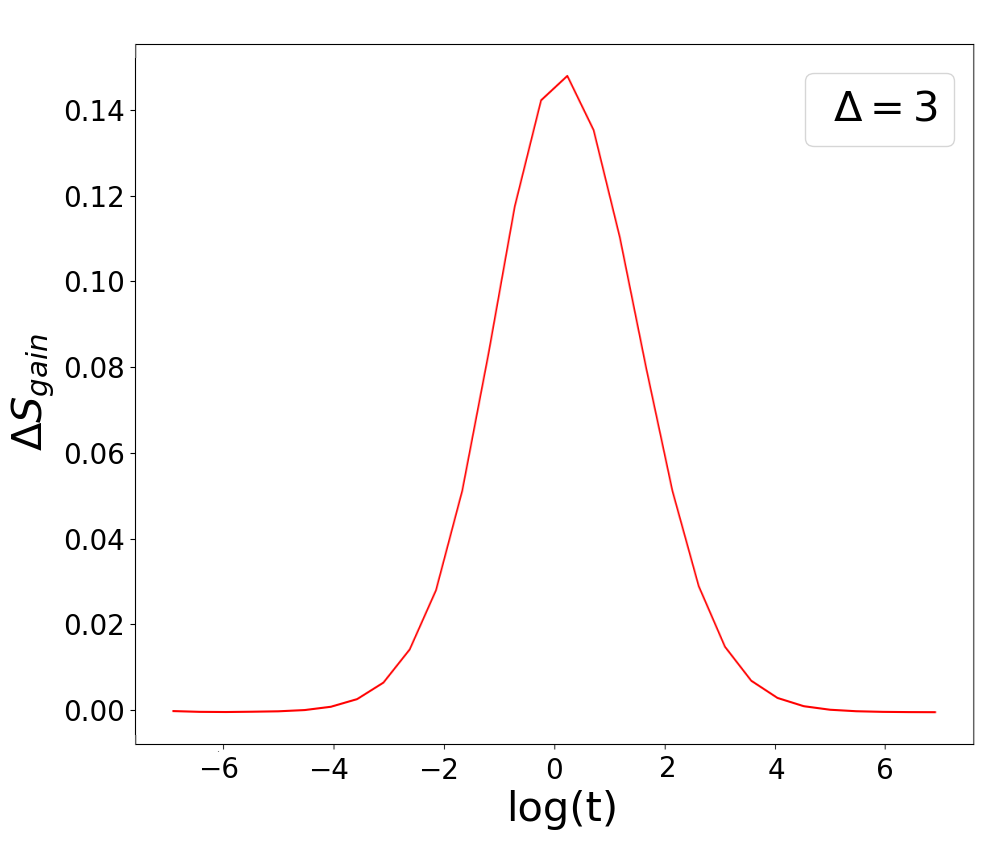}
	\caption{(a) Time dependent behavior of $\Delta S_{gain}$ for a walker in a super-cooled liquid following Ref.\cite{langer}.}
	\label{Sglassy}
\end{figure}

It was shown in \cite{langer} that the distribution for very short ($\ll \Delta$) and very long ($\gg \Delta$) times is Gaussian, but with different variances, $\Delta t$ and $2 t$ respectively. The crossover period in between is where the non-Gaussianity, measured in \cite{langer} by the quantity $\alpha_2 = \frac{3 \langle r^4 \rangle}{5 \langle r^2 \rangle^2}-1$, reaches a peak at times of the order of $\Delta^{-1}$. Fig. \ref{Sglassy} shows that our measure of non-Gaussianity, $\Delta S_{\rm{gain}}$, captures these basic properties as well.

Fig. \ref{glassywalk} (b) shows a plot of $\alpha_2$ vs. $\Delta S_{\rm{gain}}$ for $\Delta=3$. We see significant differences in the dynamic behavior between $\alpha_{2}$ and $\Delta S_{\rm{gain}}$. Time moves counter-clockwise along the loop. For low and high times (the lower left hand corner, near the origin), the behavior is mostly identical when the dynamics is essentially a Gaussian, and both $\alpha_2$ and $\Delta S_{\rm{gain}}$ are small. In the intermittent regime, however, heterogeneity is captured differently in $\alpha_{2}$ and $\Delta S_{\rm{gain}}$ as is clearly seen in Fig. 1(b).


 We now show the time dependent behavior of $\Delta S_{gain}$ for the same system in Fig. 2 for $\Delta=3.0$ that corresponds to the case in Fig. 1(b). We see $\Delta S_{gain}$ has small values at $t\rightarrow 0$ and when is $t$ sufficiently larger than the timescale associated with the heterogeneity. $\Delta S_{gain}$ has a peak near $t\approx \Delta^{-1}$. The behavior is qualitatively similar to the behavior of $\alpha_2$ reported in glassy systems\cite{cage-pc,langer,shell3,jh,guan}. However, as seen from Fig. 1 (b), the two quantities show maximum deviation near the peak. This is because for intermediate times, the heterogeneous nature of the medium strongly affects the displacement distribution $P_{ng}$. 
 At $t \gg \tau_{S}$ , $\Delta S_{\rm{gain}} = D_{KL} \left( P_{\rm{ng}} || P_{\rm{ng}}^{\rm{g}} \right) \approx D_{KL} \left( P_{\rm{ng}}^{\rm{g}}||P_{\rm{eff}} \right) \sim \ln \frac{G(x,t;D_{L})}{P_{\rm{ng}}(x,t)} \rightarrow 0$ as $P_{\rm{ng}}(x,t;D_{L}) \rightarrow P_{\rm{ng}}^{\rm{g}}(x,t;D_{L}) \rightarrow G(x,t;D_{L})$ where $D_{L}$ is the long time diffusion coefficient. This is also consistent with Figs. 1(b) and 2. In Fig. 1(b), for smaller values of $\alpha_{2}$, we observe $\Delta S_{gain}\approx \alpha_{2}$. This happens at $t\rightarrow 0$ and $t \gg \tau_{S}$ when $P_{ng}$ has a shape close to Gaussian.

The potential of information theory based frameworks in understanding thermodynamic and physical transitions has been demonstrated recently\cite{mss1,mss2,tsvi1,tsvi2}. As we have shown, the relative entropy based construction not only captures the essential factors of the non-Gaussianity, but also provides higher order information coming from higher order moments. We have shown that this information is most relevant when the heterogeneity in the system is large. For the specific model, \cite{langer} this occurs at $t \approx \Delta^{-1}$.


{\it Discussion:} Our analysis adds a new, sophisticated tool for gaining insights into heterogeneity via the role of non-Gaussianity in physical systems. The information extracted via the newly proposed relative entropy based quantification shows that the non Gaussianity inferred by moment ratios like $\alpha_2$ is not adequate in handling heterogeneous situations, as we have shown with the study of two concrete models of soft-matter systems, namely a non-Gaussian walker with a distribution of diffusivities, and a walker in a super-cooled liquid. The relative entropy integrates information from the whole distribution of walker displacements and thus provides a more complete picture than looking at the first few moments. Our study also has the potential to address the issue of modeling diffusion in heterogeneous super-cooled liquids through a distribution of diffusivities $P(D)$, since the entropic metric we introduce could be a useful tool to guide such modeling.


In brief, we extract physical information in dynamic heterogeneity when molecular motion undergoes heterogeneity in diffusion. The heterogeneity appears as a competition between intrinsic disorder due to diffusion and order arising out of spatial localization of molecular mobilities. Such interplay has been realized as a non-monotonic non-Gaussianity in the particle displacement distributions in a host of systems. In this work, we gain insights into these complex transport processes via an information-theoretic description. Out of many further possibilities, connecting microscopic heterogeneity to thermodynamics becomes an immediate natural challenge. In future, we would also investigate how the new development finds its relevance in gaining fundamental insights into the complex structural relaxation in similar systems. Our development is not at all limited to glass-forming systems, and we believe it will continue to surprise scientists from different domains where microscopic heterogeneity governs the complex transport processes and molecular metastability in the energy basin is of key interest.

\begin{acknowledgements}
We acknowledge discussions with P. Chaudhuri, S. Ghosh, C. Dasgupta and M. S. Shell. We also thank V. Vaibhav for sharing data. This research was supported in part by the International Centre for Theoretical Sciences (ICTS) during a visit for participating in the program - Entropy, Information and Order in Soft Matter (Code: ICTS/eiosm2018/08)
\end{acknowledgements}

\begin{thebibliography}{}

 
 \bibitem{dh1}
A. Einstein, Investigations on the theory of Brownian movement, Ed. R. Furth (Dover, NewYork) [ Papers I and
IV ] (1956); E. Frey and K. Kroy, Ann. Phys. {\bf 14}, 20 (2005); Jean Philibert, Diffusion fundamentals, {\bf  4}, 1 (2006)

\bibitem{sm1}
S. R. Nagel, Rev.\ Mod.\ Phys.\ {\bf 89}, 025002 (2017) 


\bibitem{ng-bouchad1}
Dynamical Heterogeneities in Glasses, Colloids, and Granular Media, Ed. L. Berthier, G. Biroli, J-P. Bouchaud, L. Cipelletti, W. van Saarloos, Oxford (2011)

\bibitem{ng-bouchad2}
J. P. Bouchaud, A. Georges, Phys. Rep., {\bf 195}, 127 (1990)

\bibitem{dh2}
W. K. Kegel, and A. van Blaaderen, Science, {\bf 287},290 (2000)


\bibitem{ng-pc}
P. Chaudhuri, L. Berthier, and W. Kob, Phys. Rev. Lett., {\bf 99}, 060604
(2007)

\bibitem{ng-granick1}
B. Wang, S.M. Antony, S.C. Bae and S. Granick, Proc. Nat. Acad. Sci.,  {\bf 106}, 15160 (2009)

\bibitem{ng-granick2}
B. Wang, J. Kuo, S.C. Bae and S. Granick, Nat. Mat., {\bf 11}, 481 (2012)

\bibitem{ng-chubensky}
M. V. Chubynsky and G. W. Slater, Phys. Rev. Lett., {\bf 113}, 098302 (2014)

\bibitem{dutta1}
S. Dutta and J. Chakrabarti, EPL, {\bf 16}, 38001(2016)

\bibitem{jain-seb}
R. Jain and K. L. Sebastian, J. Phys. Chem. B 16, 3988 (2016);  R. Jain and K. L. Sebastian, 120, 9215-9222 (2016)

\bibitem{ng-metzler1}
A. V. Chechkin, F. Seno, R. Metzler, and I. M. Sokolov, Phys. Rev. X 7, 021002 (2017)

\bibitem{ng-metzler2}
R. Metzler, Biophys J., 112(3), 413 (2017)


\bibitem{dutta3}
S. Dutta, Chem. Phys., {\bf 522}, 256 (2019)

\bibitem{cage-sk}
S. Sengupta, S. Karmakar, C. Dasgupta, and S. Sastry, J. Chem. Phys., {\bf 138}, 12A548 (2013); S. Sengupta and S. Karmakar, J. Chem. Phys., {\bf 140}, 224505 (2014)

\bibitem{cage-pc}
P. Chaudhuri, P. I. Hurtado, L. Berthier, and W. Kob, J. Chem. Phys., {\bf 142}, 174503 (2015)

\bibitem{kullback}
S. Kullback and R. A. Leibler, Ann. Math. Stat, {\bf 22}, 79 (1951).


\bibitem{shannon}
C. E. Shannon, Bell Syst. Tech. J., {\bf 27} 379 (1948).

\bibitem{iks}
J. S. Ivan, M. S. Kumar and R. Simon, Quant. Inf. Process. {\bf 11}, 853-872 (2012).

\bibitem{cover}
T. M. Cover and J. A. Thomas, 2nd Ed. ,{\it Elements of Information Theory}, John Wiley $\&$ Sons

\bibitem{langer}
J. S. Langer, $\&$ S. Mukhopadhyay (2008), PRE, 77, 061505 (2008).

\bibitem{shell3}
M. S. Shell, P. G. Debenedetti and F. H Stillinger, J. Phys.: Condens. Matter 17 (2005), S4035 (2005)

\bibitem{jh}
S. K. Schnyder, T. O. E. Skinner, A. L. Thorneywork, D. G. A. L. Aarts, J. Horbach, and R. P. A. Dullens, Phys. Rev. E 95, 032602 (2017)

\bibitem{guan}
L. Wang, N. Xu, W. H. Wang, and P. Guan, Phys. Rev. Lett. 120, 125502 (2018)

\bibitem{sk2}
B. P. Bhowmik, I. Tah, and S. Karmakar, Phys. Rev. E., {\bf 98}, 022122 (2018)

\bibitem{mss1}
M. S. Shell, J. Chem. Phys. {\bf 129}, 144108 (2008)

\bibitem{mss2}
M. S. Shell, {\it Coarse-Graining with the relative entropy}, Ed. Stuart A. Rice  Aaron R. Dinner (Wiley) (2016)

\bibitem{tsvi1}
A. Condon, H. Kirchner, D. Larivière, W. Marshall, V. Noireaux, T. Tlusty, E. Fourmentin, EMBO Rep. 19, e46628 (2018)

\bibitem{tsvi2}
J-P. Eckmann, J. Rougemont, and T. Tlusty, Rev. Mod. Phys., 91, 031001 (2019)




\end{thebibliography}

\end{document}